\documentstyle[aps,prd,twocolumn,epsf,epsfig]{revtex}
\newcommand \bea{\begin{eqnarray}}
\newcommand \eea{\end{eqnarray}}
\newcommand \ga{\raisebox{-.5ex}{$\stackrel{>}{\sim}$}}
\newcommand \la{\raisebox{-.5ex}{$\stackrel{<}{\sim}$}}
\newcommand{\av}[1]{\langle{#1}\rangle}

\begin{document}
\twocolumn[\hsize\textwidth\columnwidth\hsize
\csname@twocolumnfalse%
\endcsname
\draft
\title{Shell structure and pairing for interacting fermions in a trap}
\author{Henning Heiselberg and Ben Mottelson}
\address{NORDITA, Blegdamsvej 17, DK-2100 Copenhagen \O, Denmark}
\maketitle

\begin{abstract}
The shell structures for weakly interacting fermions in harmonic
oscillator traps at zero temperature
undergo several transitions depending on the number
of particles in the trap and their interaction strength. Calculations
of the one and two particle spectra give the pairing gaps, and the
many particle state is constructed by the seniority scheme. For
sufficiently few particles, $N\la 10^4$, the pairing field exceeds the
mean-field splitting of single particle levels at the Fermi surface
and the greater number of almost degenerate states due to the $SU(3)$
symmetry in the harmonic oscillator shell lead to {\it supergaps}.
Deformation and rotation are also discussed.
\end{abstract}
\vskip1pc]


Dilute systems of cold degenerate fermionic atoms have been trapped
and cooled down to a fraction of the Fermi temperature \cite{JILA}.
Collective modes are studied and the superfluid gap is actively
searched for.  At low particle densities the shell structures in traps
are pronounced as they are in nuclei and the level degeneracies are
important for the size of the gaps which can differ substantially from
those known from homogeneous systems \cite{gap} and systems with
continuous level densities.  The atomic traps have the advantage as
compared to, e.g., nuclei, that one can study almost any number of
particles and vary their density, interaction strength (by tuning
Feshbach resonances \cite{Feshbach}), and the number of spin states,
e.g. 18 for $^{40}$K atoms.  The object of this paper is to study the
novel shell structures of such dilute fermionic systems and to
determine the number of particles and interaction strength, at which
the larger $SU(3)$ symmetry and degeneracy of the harmonic oscillator
potential leads to enhanced gaps.


We treat a gas of $N$ fermionic atoms of mass $m$ in a 
harmonic oscillator (h.o.) potential at zero temperature interacting
via a two-body interaction with attractive s-wave scattering length $a<0$. 
We shall mainly discuss a spherically symmetric trap and a dilute gas
(i.e. where the density $\rho$ obeys $\rho|a|^3\ll 1$) of particles
with two spin states with equal population. The Hamiltonian is then given by
\bea
  H &=& \sum_{i=1}^{N} \left( \frac{{\bf p}_i^2}{2m} +
    \frac{1}{2} m\omega^2 {\bf r}_i^2 \right)
  + 4\pi \frac{\hbar^2a}{m} \sum_{i<j} 
     \delta^3({\bf r}_i-{\bf r}_{j}) \,.
\eea
For a large number of particles $N$ at zero temperature the Fermi energy is
for a non-interacting system
\bea
   E_F=(n_F+3/2)\hbar\omega \,
  \simeq \, \left(3N\right)^{1/3} \hbar\omega   \,, \label{EF}
\eea
where $n_F$ is the h.o. quantum number at the Fermi surface.
The h.o. levels are highly degenerate with states having angular momenta
$l=n_F,n_F-2,...,1$ or 0 due to the     
$U(3)$ symmetry of the 3D spherically symmetric h.o. potential.
However, interactions split this degeneracy.
In the Thomas-Fermi (TF) approximation the 
mean-field potential is
\bea
    U(r) &=& 2\pi\frac{\hbar^2a}{m} \rho(r) \,, \label{U}
\eea
the Fermi energy 
$E_F=\hbar^2k_F^2(r)/2m+(1/2)m\omega^2r^2+U(r)$,
and the density 
\bea
     \rho(r) = k_F^3(r)/3\pi^2 \,
     \simeq \rho_0 \left(1-r^2/R_{TF}^2\right)^{3/2} \,, \label{rho} 
\eea
inside the cloud $r\le R_{TF}=a_{osc}\sqrt{2n_F+3}$, where
$a_{osc}=\sqrt{\hbar/m\omega}$ is the oscillator length, and
$\rho_0=(2n_F)^{3/2}/3\pi^2a_{osc}^3$ the central density \cite{b}. 
These equations constitute the 
energy density functionals which now can be solved self-consistently. 
The latter expression in Eq.~(\ref{rho}) is valid for dilute systems
where the interactions contribute a mean-field weak compared to the
confining potential.

 The system is {\it dilute} when the Fermi energy is large compared to
the mean-field energy, $E_F\gg |U|$, or equivalently $k_F|a|\simeq
\rho^{1/3}|a|\simeq n_F^{1/2}|a|/a_{osc}\ll 1$.  {\it Dense}
Fermi and Bose systems are studied in Refs. \cite{HH} and
\cite{Bose}.  We shall be particularly interested in {\it very dilute}
systems where also the h.o.  energy exceeds the mean-field
potentials, $\hbar\omega\gg |U|$, or equivalently
$n_F^{3/2}|a|/a_{osc}\ll 1$ (see Fig. 3). 

The splitting of the degenerate $l=n_F,n_F-2,...,1$ or 0 states
by the mean-field potential 
can be calculated perturbatively in the very dilute limit.
From
the radial h.o. wave function ${\cal R}_{nl}(r)$ for the state
with angular momentum $l$ and $n$ radial nodes in the h.o. shell
$n_F=2n+l$, we obtain the single particle energies
\bea 
  \epsilon_{n_F,l}-\left(n_F+\frac{3}{2}\right)\hbar\omega &=& \int U(r) 
 |{\cal R}_{nl}(r)|^2 r^2dr \nonumber\\
  &=& \frac{4\sqrt{2}}{3\pi}\frac{a}{a_{osc}} n_F^{3/2} \hbar\omega\, F(n_F,l)
 \,, \label{EMF}
\eea
where the mean-field and wave function overlap is
\bea
  F(n_F,l)&=& \int_0^{R_{TF}}
 |{\cal R}_{nl}(r)|^2\left(1-\frac{r^2}{R_{TF}^2}\right)^{3/2}  
              r^2dr \label{I}\\
   &\simeq&\frac{4}{3\pi}-\frac{1}{4\pi} \frac{l(l+1)}{n_F^2} +O(n_F^{-1}) \,.
 \label{IWKB} 
\eea
The latter result is obtained using the WKB approximation \cite{WKB}
which compares well to numerical results for all 
$l$ when $n_F\gg 1$ as shown in Fig.~1. 
The mean-field energies are proportional to the coupling and $n_F^{3/2}$
and are split like rotational bands with a prefactor that
is smaller than the average mean-field
energy by a factor $\sim3/16$. The relative small splitting
reflects the fact that the mean-field potential is almost
quadratic in $r$  and the anharmonic terms therefore small.
For attractive interactions ($a<0$)
the lowest-lying states have small angular momentum, which is {\it opposite}
to nuclei.

\vspace{-0.5cm}
\begin{figure}
\begin{center}
\psfig{file=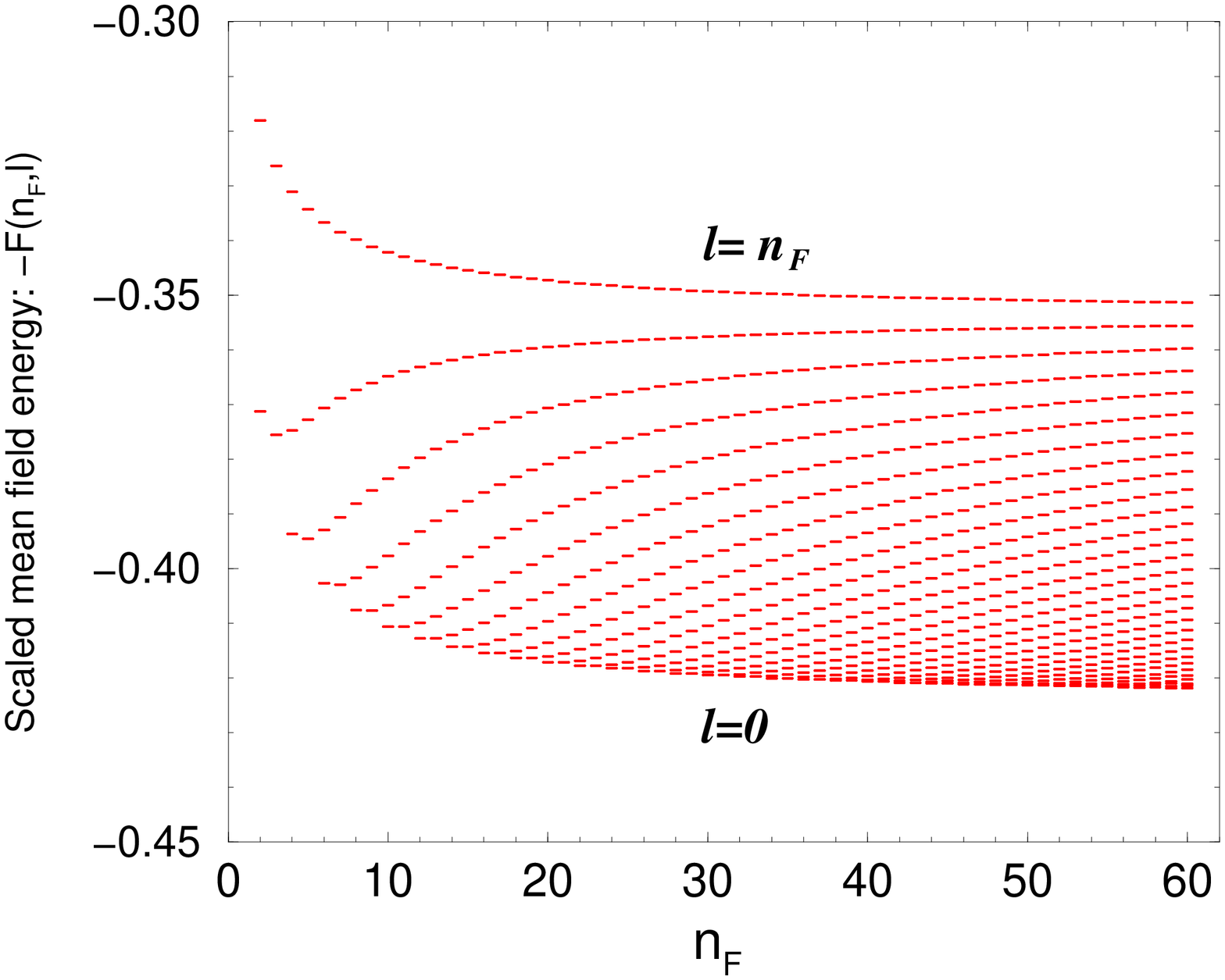,height=8.0cm,angle=-90}
\vspace{.2cm}
\begin{caption}
{Single particle levels for very dilute systems of fermions in
h.o. shells $n_F=2n+l$,
The mean-field splitting $F(n_F,l)$ vs. $n_F$
of multiplets $l=n_F,n_F-2,...,1$ or 0 are
calculated numerically from Eq. (\ref{I}).}
\end{caption}
\end{center}
\label{MFlevels}
\end{figure}


The wave function for a pair of particles each with angular momentum $l$
adding up to total angular momentum $L$ 
is a sum of time-reversed two-particle states
\bea
   \phi_{L}(n_F,l) = \sum_m \av{lml-m|L0} 
 \psi_{lm}({\bf r}_1)\psi_{l-m}({\bf r}_2) \label{wf}\,,
\eea
for $M=0$.
Here $\psi_{lm}({\bf r})={\cal R}_{nl}(r)Y_{lm}(\theta,\phi)$ 
is the single particle wave function normalized to unity and $m=-l,...,l$. 
As the spin wave function is always in a singlet state, 
since otherwise the interactions vanish, it has been excluded in 
Eq.\ (\ref{wf}). The corresponding gap equal to 
half the pairing energy between two particles, is
\bea
&\Delta_L(n_F,l)&\, =\,\frac{1}{2}\av{\phi_{L}(n_F,l)|4\pi\frac{\hbar^2|a|}{m}
     \delta^3({\bf r}_1-{\bf r}_2)|\phi_{L}(n_F,l)}     \nonumber\\
  &=& \frac{\hbar^2|a|}{2m}(2l+1)^2  \frac{\av{l0l0|L0}^2}{2L+1} 
  \int|{\cal R}_{nl}(r)|^4r^2dr  \,,\label{Del}
\eea
for even $L$ and zero otherwise.
The gap is largest when pairs of particles are
coupled to total angular moment $L=0$
for which the Clebsch-Gordan coefficient is $\av{l0l0|00}^2=(2l+1)^{-1}$.

The overlap integrals can be calculated for $n_F\gg 1$
within the WKB approximation \cite{WKB}
\bea
  \frac{\Delta_0(n_F,l)}{(|a|/a_{osc})\hbar\omega} &=&
    \left(  \begin{array}{ll}
    \frac{1}{\sqrt{\pi}} & , \, l=n_F \\
    \frac{\sqrt{2}}{\pi^2\sqrt{n_F}} {\cal L} & , \, 1\ll l, 1\ll n_F-l \\
    \frac{\sqrt{2}}{\pi\sqrt{n_F}}& , \, l=0  \end{array} \right)
   .          \label{Gll}
\eea
Here, the function ${\cal L}(n_F,l)\simeq \ln(l)$, 
to leading logarithmic accuracy.
Numerical calculations of the gaps are shown in Fig. 2.

\vspace{-0.cm}
\begin{figure}
\begin{center}
\psfig{file=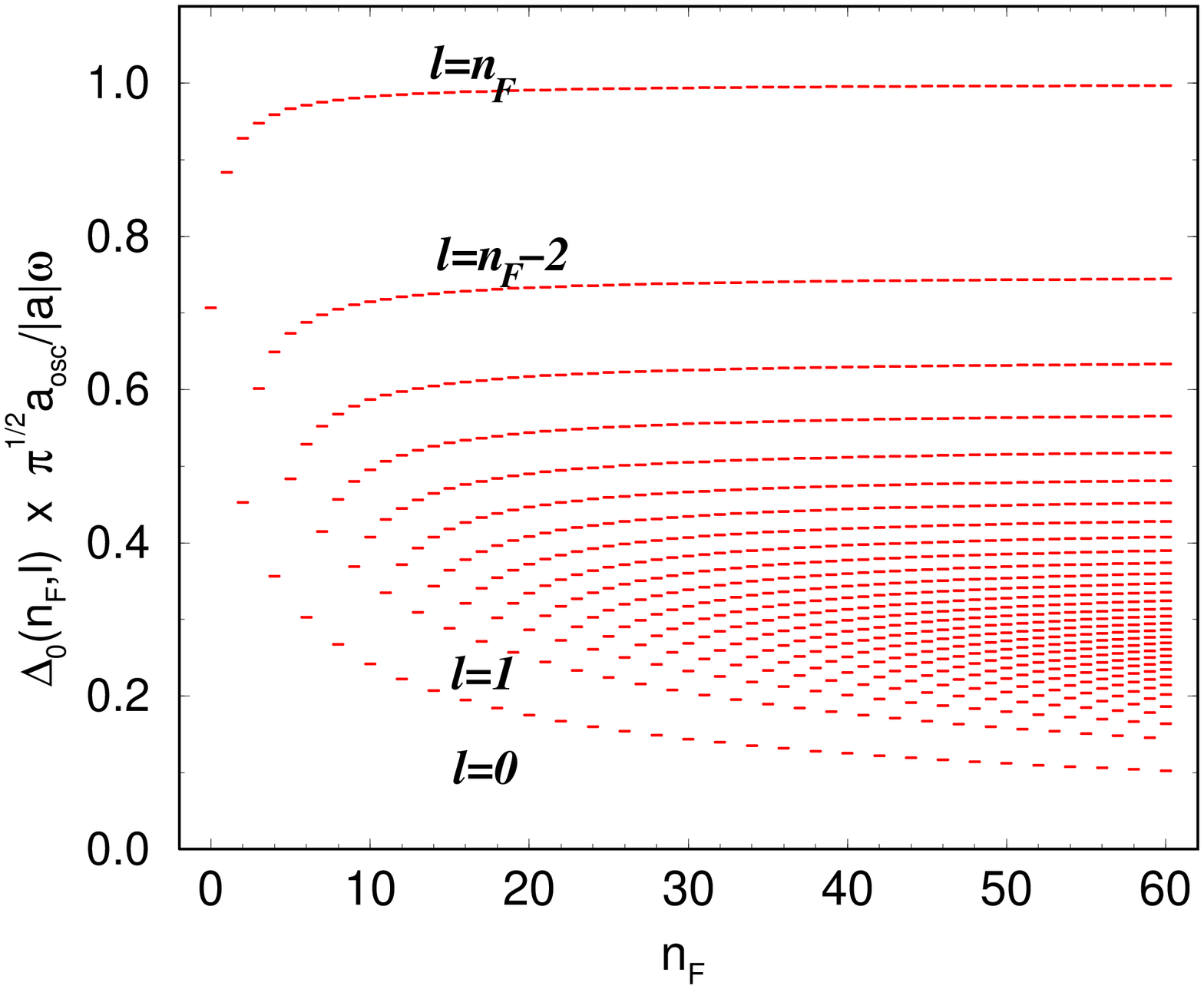,height=8.0cm,angle=-90}
\vspace{.2cm}
\begin{caption}
{Pairing energies $\Delta_0(n_F,l)$ in units of 
$\pi^{-1/2}\hbar\omega|a|/a_{osc}$ calculated from 
Eq. (\protect\ref{Del}) (see also Eq. (\protect\ref{Gll})).
For given $n_F$ the levels are $l=n_F,n_F-2,...,1$ or 0 from top to bottom.}
\label{del}
\end{caption}
\end{center}
\end{figure}

For two or more particles in a shell,
it is known from nuclear pairing theory that 
it is generally  energetically favorable
for the particles to pair in states with $L=0$ where their
spatial overlap is maximum. In the degenerate pairing model
with interactions only in the $L=0$ pair state
the pairing energy between $n_v$ particles in a shell is
\bea
  \Delta(n_v,v) = -\,\frac{(n_v-v)(2\Omega-n_v-v+2)}{4\Omega} \Delta_0 \,,
 \label{sen}
\eea
where the {\it seniority} $v$ is the number of unpaired particles.
The number of available states in the multiplet is $\Omega=2(2l+1)$ 
for the multiplet $l$
with pairing gaps $\Delta_0=\Delta_0(n_F,l)$. 
In the ground state for even $n_v$ all particles are paired
i.e. $v=0$, and the lowest excitations are reached by breaking pairs
\cite{BM}.

When $n_F$ (or $l$) is sufficiently small, however, the mean-field
splitting of Eq.\ (\ref{EMF}) is smaller than the pairing gaps and the
pairing acts between all states in an oscillator shell and not just in a
single $l$ multiplet, i.e. the full $SU(3)$ symmetry is effectively
restored as compared to the $SU(2)$ symmetry of the single $l$
multiplet. We shall refer to this enhanced pairing as
``$SU(3)$-pairing'' or ``super-pairing''. 
We shall calculate the gap first by writing down the ``super''-pair
wave function and secondly by solving the gap equation.

Assuming $SU(3)$ symmetry the
pairing can be calculated variationally
with a  pair wave function that is a generalization of  Eq. (\ref{wf}) 
to a sum over $l=n_F,n_F-2,....,1$ or 0
\bea
   \Phi_0(n_F) = 
    \frac{\sum_l \alpha_l\, \phi_0(n_F,l)}
     {[\sum_l \alpha_l^2]^{1/2}} \,.
\eea
The weights $\alpha_l$ can be found by  a variational method
with the overlap integrals
$\int{\cal R}_{nl}^2{\cal R}_{n'l'}^2r^2dr$ calculated numerically.
For large $n_F$ we find $\alpha_l\sim\sqrt{2l+1}$ very accurately.
The sums then reduce to the h.o. level density:
$\sum_{lm}|\psi_{lm}({\bf r})|^2=(4\pi)^{-1}\sum_l (2l+1){\cal R}_{nl}^2 
=(1/2)d\rho(r)/dn_F$, and the supergap becomes
\bea
  G &=& \frac{1}{2}\av{\Phi_0(n_F)|4\pi\frac{\hbar^2|a|}{m}
        \delta^{3}({\bf r}_1-{\bf r}_2)  |\Phi_0(n_F)} \nonumber\\
   &\simeq& \pi\frac{\hbar^2|a|}{m}
     \frac{ \int [d\rho(r)/dn_F]^2 d^3r}{\int [d\rho(r)/dn_F] d^3r} \label{G} \\
  &=& \frac{32\sqrt{2}}{15\pi^2} n_F^{1/2}
      \frac{|a|}{a_{osc}} \hbar\omega \,,  \label{Gr}
\eea
with the particle density $\rho(r)$ from Eq. (\ref{rho}).
The gap is a factor $n_F$ larger than that for a single $l$ value
because the level degeneracy is larger by $n_F$.

Turning the mean-field splitting back on,
the condition for superpairing is that the pairing field $2G$
exceeds the mean-field splitting of the single
particle levels, $\epsilon_{n_F,l=n_F}-\epsilon_{n_F,l=0}$.
Using Eqs.\ (\ref{EMF}), (\ref{IWKB}) and (\ref{Gr}) 
the condition for superpairing becomes
\bea
  n_F \, \la \, \frac{64}{5} \,,
\eea
corresponding to $N\la 10^3$ trapped particles.
For $n_F>64/5$ we find that only the lower lying states 
$l\la 64/5$ superpair.
The seniority scheme of Eq. (\ref{sen}) also applies to many superpairs 
replacing $\Delta_0=G$ and the level degeneracy
for a full h.o. shell $\Omega=(n_F+1)(n_F+2)$.

The transition from pairing within a single $(n_F,l)$ level to many levels
can be calculated by solving the gap equation
\bea
\Delta(n_F,l) &=& -\frac{1}{2}\, \sum_{l'} 
  \frac{\Delta(n_F',l')}{\sqrt{\Delta^2(n_F',l')+(\epsilon_{n_F,l'}-\mu)^2}}
  \sqrt{\frac{2l'+1}{2l+1}}  \nonumber \\ 
 &&
   \av{\phi_0(n_F',l')|4\pi\frac{\hbar^2a}{m}\delta^3({\bf r}_1-{\bf r}_2)|
      \phi_0(n_F,l)}  \,. \label{GE}
\eea
within the same shell, i.e., $n_F'=n_F$.
Here, the single particle energies are given by Eq. (\ref{EMF})
in the very dilute limit and the number of particles is such that
the Fermi level is positioned at
$\mu=\epsilon_{n_F,l}$. 
When $n_F\la 64/5$ the gap equation reproduces Eq. (\ref{G}),
which is followed by a transition region as shown in Fig. 3.
When $n_F\ga (64/5)\ln(n_F)$, equivalent to $n_F\ga50$, the pairing gap
approach that for a single $l$ level as given by Eq. (\ref{Gll}).

When the interaction becomes sufficiently strong pairing also takes
place between different h.o. shells.  The pair wave function can still
be expressed as a sum over time-reversed single particle wave
functions \cite{BB}.  For interaction strengths
$G\ln(n_F)\la\hbar\omega$ Eq. (\ref{wf}) is still a good approximation
and the generalized gap equation is then simply Eq.\ (\ref{GE}), but
also summed over $n_F'\ne n_F$ \cite{BH}.  For strong interactions,
$2G\ln(n_F)\ga\hbar\omega$, the radial wave functions differ
substantially from the h.o. ones due to the strong pair field
\cite{Baranov,BH}.  The coherence length $\xi\simeq
R_{TF}\hbar\omega/\Delta$ is smaller than the system size and bulk
superfluidity sets in (see Fig. 3).  The finite system behaves like an
irrotational fluid and exhibits collective oscillations with
frequencies of order $\sim\hbar\omega$ \cite{Bruun}.

\vspace{0.cm}
\begin{figure}
\begin{center}
\psfig{file=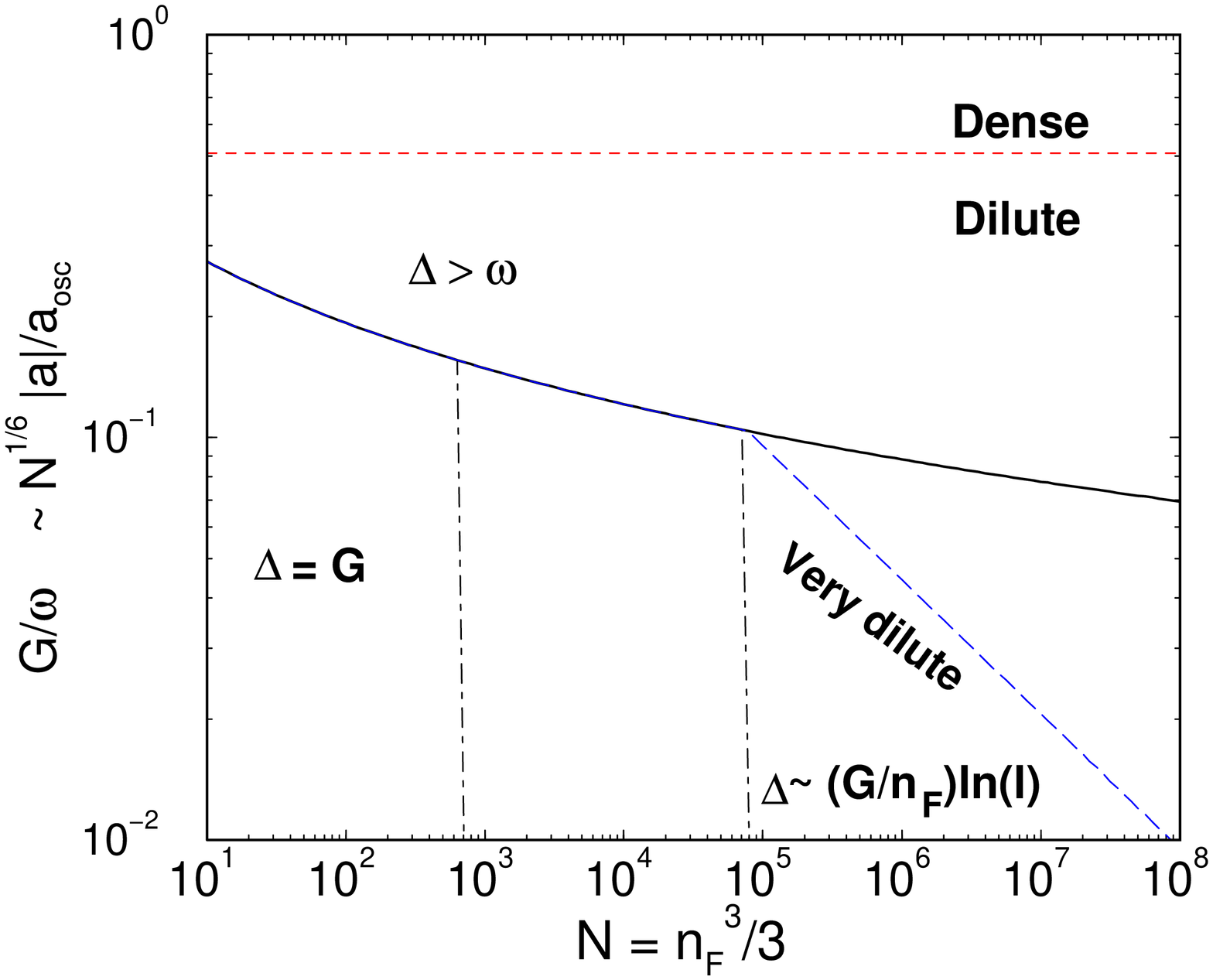,height=8.0cm,angle=-90}
\vspace{.2cm}
\begin{caption}
{Diagram displaying the transitions of the level structures and pairing gaps 
$\Delta$ (in units of $\hbar\omega$) in traps as a function of 
the number of particles $N=n_F^3/3$
and the scaled interaction strength 
$G/\hbar\omega=(32/15\pi^2)(2n_F)^{1/2}|a|/a_{osc}$.
The dashed lines separate the dense, dilute and very dilute trapped gases.
The region between the dash-dotted lines indicates the transition between
superpairing, $\Delta=G$, and pairing in a single $l$-level, 
$\Delta\simeq(G/n_F)\ln(l)$.
Above the full line $G=\hbar\omega/2\ln(2n_F)$ 
inter-shell pairing enhances the gap $\Delta$
above $\hbar\omega$}
\end{caption}
\end{center}
\label{phase}
\end{figure}

Whereas nuclei with partially filled shells generally are quite deformed, 
$\delta=\langle r_3^2\rangle/\langle r_1^2\rangle-1\sim 1/n_F$ \cite{BM}, 
the atomic mean-field does not
deform in bulk in a spherically symmetric h.o. trap. 
The h.o. shell at the Fermi surface does not deform either
because the energy of the superfluid state is always lower than that
of the deformed state, and the atoms will be a spherically symmetric
fluid.
  
Deformed traps such as the oblate ones with
$\omega_\perp\equiv\omega_1=\omega_2\la\omega_3$ will only
have $SU(2)$ symmetry and correspondingly lower level degeneracies
except when ratios of the trap frequencies are rational numbers.
The Fermi energy for the deformed h.o. potential is $E_F^0\simeq
n_F\hbar\omega_\perp + n_{F,3}\hbar\omega_3$, where $n_F$ now is the
h.o. quantum number in two dimensions which has degenerate single
particle states of angular momentum projected along the symmetry axis:
$m=-n_F,-n_F+2,...,n_F$.

For very deformed traps such that 
$\omega_\perp\ll\omega_3N^{-1/2}$, the h.o.
potential is effectively two-dimensional, 
$U(r)\propto (1-r^2/R_{TF}^2)\theta(R_{TF}-r)$, inside the cloud. 
Such a quadratic mean-field
potential does not destroy the $SU(2)$ symmetry or split the h.o. shells,
except for the states with lowest angular momentum whose wave functions
extend slightly outside $R_{TF}$.
The pairing gaps can be calculated
for two particles in shell $n_F$ \cite{Magicref} exploiting the SO(2,1)
symmetry \cite{Pit} of a two-dimensional trap for particles with
delta-function interactions
\bea
 \Delta^{2D}(S,M) &=& \frac{1}{2}\av{SM|4\pi\frac{\hbar^2|a|}{m}
 \delta^3({\bf r}_1-{\bf r}_2)|SM}   \nonumber\\
     &=& \frac{|a|}{a_3}\hbar\omega_\perp \,\frac{1}{2^{2S}} \,
 \frac{(S-M)!(S+M)!}{[(\frac{S-M}{2})!(\frac{S+M}{2})!]^2} \,,\label{magic}
\eea
where $M=-S,-S+2,....,S-2,S$, is half the total angular momentum projected
along the symmetry axis,
and $S=0,1,2,...,n_F$, is a quantum number similar to seniority.
The spatial extent of the particles along the symmetry axis enters 
through 
$a_3^{-1}=\int|\psi_{n_{F,3}}(r_3)|^4dr_3=(m\omega_3/2\pi\hbar)^{1/2}$.
It is a remarkable feature that
the pairing energies are {\it independent} of $n_F$ in two dimensions. 
In the supershell $n_F$, pair energies of the states with $S\le n_F-1$ equal
those of the previous supershell $n_F-1$, recursively.
Thus, for example, the energy spectrum corresponding to 
Eq. (\ref{magic}) always has states
with excitation $2\hbar\omega_\perp$, which follows from the SO(2,1)
symmetry \cite{Pit} also for strong interactions.
For more than two particles in a shell the low lying excitations will be
dominated by paired states with the seniority quantum number
describing the number of unpaired particles.

We now turn to experimental observation of superfluidity and gaps
which requires temperatures below $T\la\Delta/k_B$.  For temperatures
$T\sim 0.2E_F$ as reached in \cite{JILA} this would require strongly
interacting atoms (e.g. near a Feshbach resonance) with inter-shell
pairing so that $\hbar\omega\la\Delta\la E_F=n_F\hbar\omega$.
Alternatively, a cloud of $\sim10^3$ $^6Li$ atoms, which have a very
large and negative natural scattering length, the supergap is
$\Delta=G\simeq 10^{-2}E_F$ and cooling below that would be required
in order to measure the gap.

Superfluid gaps can be determined in experiments with cold trapped
Fermions from response functions for multipole modes \cite{Bruun} or
from the scissors mode \cite{Marago} in a deformed trap.
Alternatively, we suggest rotating the deformed trap slowly around an
axis perpendicular to the symmetry axis. When $\Delta\ga
\delta\hbar\omega$ superfluidity reduces the moment of inertia from
the rigid-body one to that of an irrotational fluid, which is smaller by a
factor $\delta^2$ for small deformations. As for rapidly rotating
nuclei \cite{BM} several interesting phenomena such as
superdeformation, backbending, fission, etc. may occur in deformed
traps.  Optical lattices in current experiments \cite{optical} have
few atoms in each local trap and we thus expect superpairing which
favors the insulator vs. conductor state.

 In summary, we have calculated the shell structure for a dilute
system of interacting fermions as a function of the 
interaction strength and number of particles in a h.o. trap.  For weak
interactions, $G\la\hbar\omega/2\ln(n_F)$,
and few particles $N\la 10^3$ in the trap, the $U(3)$
symmetry of the h.o. leads to larger degeneracy and superpairing occurs
with supergaps $\Delta=G$.  For a larger number of particles
$N\ga10^4-10^5$, the mean-field energy generally dominates and splits
the $U(3)$ symmetry of the h.o. into multiplets by an energy that is
large compared with the gaps.  For two dimensional traps, however, the
mean-field does not split the $U(2)$ degeneracy and the gaps display a
remarkable regularity as given by Eq. (\ref{magic}).  
Atoms with more
than two spins have larger level degeneracies for spin independent
interactions, and the gaps are correspondingly larger as is also the
case for uniform density \cite{gap,HH}.

We thank G. Bruun and C.~J.~Pethick for valuable discussions.


\begin{thebibliography}{99}
\bibitem{b} In nuclear physics constant central density is required
and so $a_{osc}\to b\sim N^{1/6}\rho_0^{-1/3}$.
\bibitem{WKB} For $n\gg 1$ the WKB approximation
can be applied: ${\cal R}_{nl}(r)\propto \sin(k_F(r)r)/k_F(r)^{1/2}r$,
between the turning points
$r^2_\pm=(n_F+3/2\pm\sqrt{(n_F+3/2)^2-l(l+1)})a_{osc}^2$.
When the wave function has many nodes $1\ll l\ll n_F$, 
the oscillations can be averaged $\av{\sin^2(kr)}=1/2$ and 
$\av{\sin^4(kr)}=3/8$. Near the turning points the wave functions 
must be replaced by Airy functions for detailed gap calculations.
\bibitem{Magicref} The technical proof will be presented elsewhere.
\end{thebibliography}
\end{document}